\documentclass[]{elsarticle}

\usepackage[english]{babel}
\usepackage{cleveref}
\usepackage{dcolumn}
\usepackage{color}
\usepackage{hyperref}
\usepackage{bm}
\usepackage{xr}

\usepackage{subcaption}
\usepackage{amsmath}
\usepackage{amssymb}
\usepackage{graphicx}
\usepackage{subcaption}
\usepackage{color}
\usepackage{tabularx}
\usepackage{tikz,pgfplots,tikz-3dplot}
\usepackage{bbm}
\usepackage{multirow}
\usepackage{mathtools, nccmath}

\usepackage{tikzscale}
\usetikzlibrary{arrows}
\usetikzlibrary{snakes}
\usetikzlibrary{shapes}
\usetikzlibrary{shadows}
\usetikzlibrary{trees}
\usetikzlibrary{arrows}
\usetikzlibrary{positioning}                
\usetikzlibrary{calc,through}               
\usetikzlibrary{decorations.pathreplacing}  
\usetikzlibrary{calc}
\usetikzlibrary{fadings}
\usepackage{tabularx}

\usepackage[mathlines, displaymath, running]{lineno}

\newcommand*\patchAmsMathEnvironmentForLineno[1]{%
	\expandafter\let\csname old#1\expandafter\endcsname\csname #1\endcsname
	\expandafter\let\csname oldend#1\expandafter\endcsname\csname end#1\endcsname
	\renewenvironment{#1}%
	{\linenomath\csname old#1\endcsname}%
	{\csname oldend#1\endcsname\endlinenomath}}%
\newcommand*\patchBothAmsMathEnvironmentsForLineno[1]{%
	\patchAmsMathEnvironmentForLineno{#1}%
	\patchAmsMathEnvironmentForLineno{#1*}}%
\AtBeginDocument{%
	\patchBothAmsMathEnvironmentsForLineno{equation}%
	\patchBothAmsMathEnvironmentsForLineno{align}%
	\patchBothAmsMathEnvironmentsForLineno{flalign}%
	\patchBothAmsMathEnvironmentsForLineno{alignat}%
	\patchBothAmsMathEnvironmentsForLineno{gather}%
	\patchBothAmsMathEnvironmentsForLineno{multline}%
}

\newcommand{\CC}{C\nolinebreak\hspace{-.05em}\raisebox{.4ex}{\tiny\bf +}\nolinebreak\hspace{-.10em}\raisebox{.4ex}{\tiny\bf +}}
\def\CC{{C\nolinebreak[4]\hspace{-.05em}\raisebox{.4ex}{\tiny\bf ++}}}

\newcommand {\e}[1]{\mathrm{~#1}}
\newcommand {\ee}[1]{\mathrm{#1}}
\modulolinenumbers[2]

\journal{Elsevier Science}

\begin{document}

\begin{frontmatter}

\title{Global Decay Chain Vertex Fitting at $B$-Factories}

\author[A]{J.-F.~Krohn\tnoteref{correspondingauthor}}
\author[A]{P.~Urquijo}
\author[D]{F.~Abudin{\'e}n}
\author[B]{S.~Cunliffe}
\author[B]{T.~Ferber}
\author[C]{M.~Gelb}
\author[C]{J.~Gemmler}
\author[C]{P.~Goldenzweig}
\author[C]{T.~Keck}
\author[B]{I.~Komarov}
\author[E]{T.~Kuhr}
\author[D]{L.~Ligioi}
\author[F]{M.~Lubej}
\author[G]{F.~Meier}
\author[C]{F.~Metzner}
\author[C]{C.~Pulvermacher}
\author[E]{M.~Ritter}
\author[H]{U.~Tamponi}
\author[B]{F.~Tenchini}
\author[F]{A.~Zupanc}
\tnotetext[correspondingauthor]{Corresponding author.\\ Email: jkrohn@student.unimelb.edu.au}

\address[A]{University of Melbourne, Melbourne, Australia}
\address[B]{Deutsches Elektronen-Synchrotron, Hamburg, Germany}
\address[C]{Karlsruhe Institute of Technology, Karlsruhe, Germany}
\address[D]{Max-Planck-Institut f\"ur Physik, Munich, Germany}
\address[E]{Ludwig Maximilians Universit\"at, Munich, Germany}
\address[F]{Jo\v{z}ef Stefan Institute, Ljubljana, Slovenia}
\address[G]{University of Sydney, Sydney, Australia}
\address[H]{INFN - Sezione di Torino, Torino, Italy}
\address{Belle~II analysis software group}

\begin{abstract}
We present a particle vertex fitting method designed for $B$~factories. The presented method uses a Kalman Filter to solve a least squares estimate to globally fit decay chains, as opposed to traditional methods that fit each vertex at a time. It allows for the extraction of particle momenta, energies, vertex positions and flight lengths, as well as the uncertainty
estimates of these quantities. Furthermore, it allows for the precise extraction of vertex parameters in complex decay chains containing neutral final state particles, such as $\gamma$ or $K_L^0$, which cannot properly be tracked due to limited spatial resolution of longitudinally segmented single-layer crystal calorimeters like the Belle~II ECL.  The presented technique can be used to suppress combinatorial background and improve resolutions on measured parameters. We present studies using Monte Carlo simulations of collisions in the Belle~II experiment, where modes with neutrals are crucial to the physics analysis program.
\end{abstract}

\end{frontmatter}

\linenumbers

	\section{Introduction} 
Particle vertex fitting techniques are widely used in particle and nuclear physics. Beyond the suppression of background, applications range from the improvement of particle momentum resolution (under the assumption they originate from some vertex point), to the determination of the presence of intermediate particles and the precision determination of decay vertex positions. One can, for example, combine the measurements of two charged pion tracks originating from the decay of a $K^0_{\rm{S}}$ to extract the decay vertex position, flight length and four-vector and their uncertainties. By performing a kinematic fit, one obtains an improvement of the pion track momenta and can use the $\chi^2$ probability of the fit result to suppress background. 

\begin{figure}
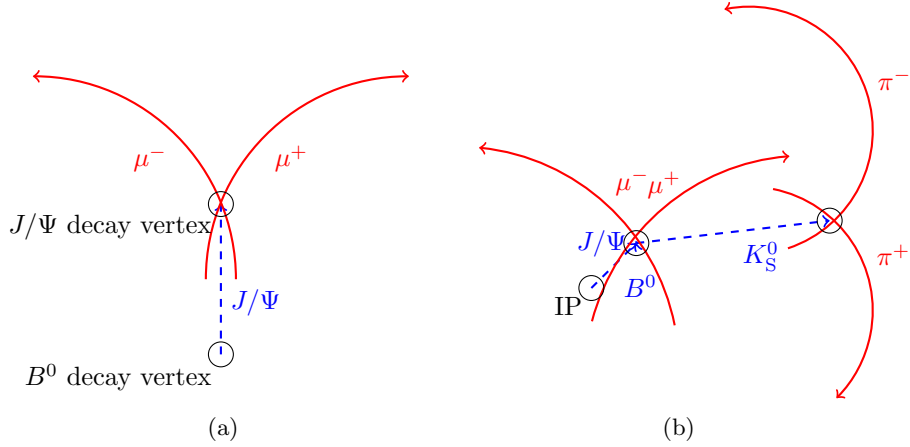
\centering
	\begin{subfigure}[t]{.49\textwidth}
		\includegraphics{tikz/jpsi.tikz}
		\caption{}
			\end{subfigure}
	\begin{subfigure}[t]{.49\textwidth}
		\includegraphics[width=1.1\textwidth]{tikz/jpsi_pipi.tikz}
		\caption{}
	\end{subfigure}
	\caption{a) Depiction of a $J/\Psi \to \mu^+\mu^- $ decay. The red lines show the track helix approximations obtained from the tracking detectors, the blue dashed lines show the decaying particle momentum vectors found by the fit. Since the decay length of the $J/\Psi$ is too short to be seen in the detector, its decay vertex is taken to be the one of the $B^0$.
		b) Depiction of a $B^0 \to J/\Psi(\to \mu^+\mu^-) K_{\rm{S}}^0(\to \pi^+\pi^-)$ decay. Measured track helices do not necessarily overlap in three dimensions. The depicted length ratios are not to scale.}
	\label{fig:classic}
\end{figure}

\begin{figure}\centering
	
	\includegraphics{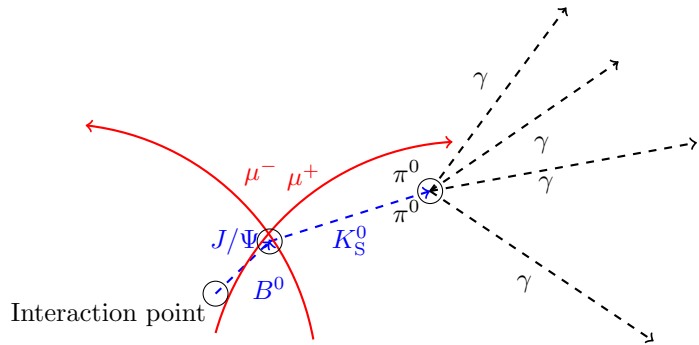}
	\caption{Depiction of a the decay $B^0 \to K^0_{\rm{S}}J/\Psi$ . The red lines show the track helix approximations obtained by the tracking detectors, the blue dashed lines show the composite particle momentum vectors found by the fit. The dashed black lines depict the photon momenta found by the fit. Note that these can only be extrapolated by the fit as the directional information of the calorimeter is not sufficient. The initial guess is that they point from the interaction point towards the calorimeter cluster. The decay lengths of the $J/\Psi $ and $\pi^0$ are too short to be seen in the detector therefore the vertex positions are taken from the particle above them in the hierarchy. 
	}
	\label{fig:classic4}
\end{figure}

In order to construct more complex decay topologies, one usually combines cascades of these fits starting with long lived stable particles such as electrons, muons, pions and photons, forming intermediate resonances and finally the full decay of interest. For example, in the decay $B^{0} \to  J/\Psi K^{0}_{\rm{S}}$, where  $J/\Psi \to \mu^{+}\mu^{-}$ and $ K^{0}_{\rm{S}}\to \pi^{+}\pi^{-}$, one would first fit the $J/\Psi$ and $K^0_{\rm{S}}$ candidates and then use these to construct the $B^0$ candidates, as depicted in Fig.~\ref{fig:classic}a and Fig.~\ref{fig:classic}b. 
However, this only works well if the final state particles are charged and leave traces in the tracking detectors. Neutral particles can not be tracked; they are only measured by their energy deposition in the calorimeter. Single layer crystal calorimeters do not offer directional information on where the particle associated to an energy deposition originated from. This means that the decay vertex can not be extracted from the fit. In order to obtain the momentum vector, it has to be assumed that the particle originates from the primary interaction point and travels directly into the cluster's center of gravity. This can introduce a large bias on the momentum components. Consider, for example, the decay $B^0\to J/\Psi K^0_{\rm{S}} $, where the kaon instead decays to neutral final states, $K^0_{\rm{S}} \to \pi^0\pi^0$, and the pions decay, $\pi^0 \to \gamma \gamma$, as displayed in Fig.~\ref{fig:classic4}.
Weakly decaying intermediate particles such as $K^0_{\rm{S}}$ have relatively long flight lengths, up to $15\e{cm}$ for $K^0_{\rm{S}}$ at Belle II. Thus for a neutral particle in such a particle decay chain, the assumption that it originates from the primary $e^+e^-$ collision is not sufficient.

The method we present in this paper overcomes these issues by globally fitting the entire decay tree in a single fit, taking into account all intermediate particles, extracting all involved particle's four-momenta, vertex positions, flight lengths and their covariance matrices, using a Kalman Filter as described in Ref.~\cite{Hulsbergen:2005pu}. We use the software environment of Belle II and rely on the \CC\  template library EIGEN~\cite{eigenweb} for matrix operations, which provides a fast execution time for the fit algorithm. We furthermore present physics applications of the fitter with Belle II Monte Carlo samples.

\subsection{The Belle II Experiment}
The algorithm described in this paper was developed for the analysis software framework of Belle II. 
The Belle II experiment takes place at the asymmetric $e^+e^-$ collider, SuperKEKB. SuperKEKB  provides a beam energy slightly above the mass of the $\Upsilon(4\rm{S})$ resonance $(10.58\e{GeV}/c^2)$ at an instantaneous luminosity of $8\cdot10^{35}\e{cm^{-2} s^{-1}}$. The $\Upsilon(4\rm{S})$ resonance  decays into pairs of $B$-mesons just above production threshold, hence this type of experiment is referred to as a $B$-factory. The asymmetric beam energy gives the $B$-meson a relativistic boost along a direction close to the detector's axis of symmetry, increasing its flight length in the lab frame, which makes it easier to study the time evolution of $B$ decays - a key observable in the study of CP symmetry violation.

The Belle II detector has a cylindrical structure designed to study the decays of $B$- and $D$-mesons, $\tau$-leptons and other processes produced in $e^+e^-$ collisions. Six layers of silicon vertex detectors (2 layers of silicon pixels (PXD), and 4 layers of double sided silicon detectors (SVD)) are located in the central volume of the detector, designed to accurately track the flight paths of charged particles. The following layers are, a central drift chamber (CDC) used to measure track trajectories within a solenoid magnetic field, Cherenkov light based particle identification devices surrounding the CDC in the barrel (TOP) and forward regions (ARICH), followed by the CsI(Tl) electromagnetic calorimeter (ECL). The outermost layers are composed of, a magnet solenoid, a $K_{\rm{L}}^0$ and muon detector system (KLM), which is also used as the flux return yoke of the magnet. The magnetic field is aligned along the detector's axis of symmetry. For a more detailed description of the detector see Ref.~\cite{techdesign}.

\begin{figure}
			\centering
		\includegraphics[width=\linewidth]{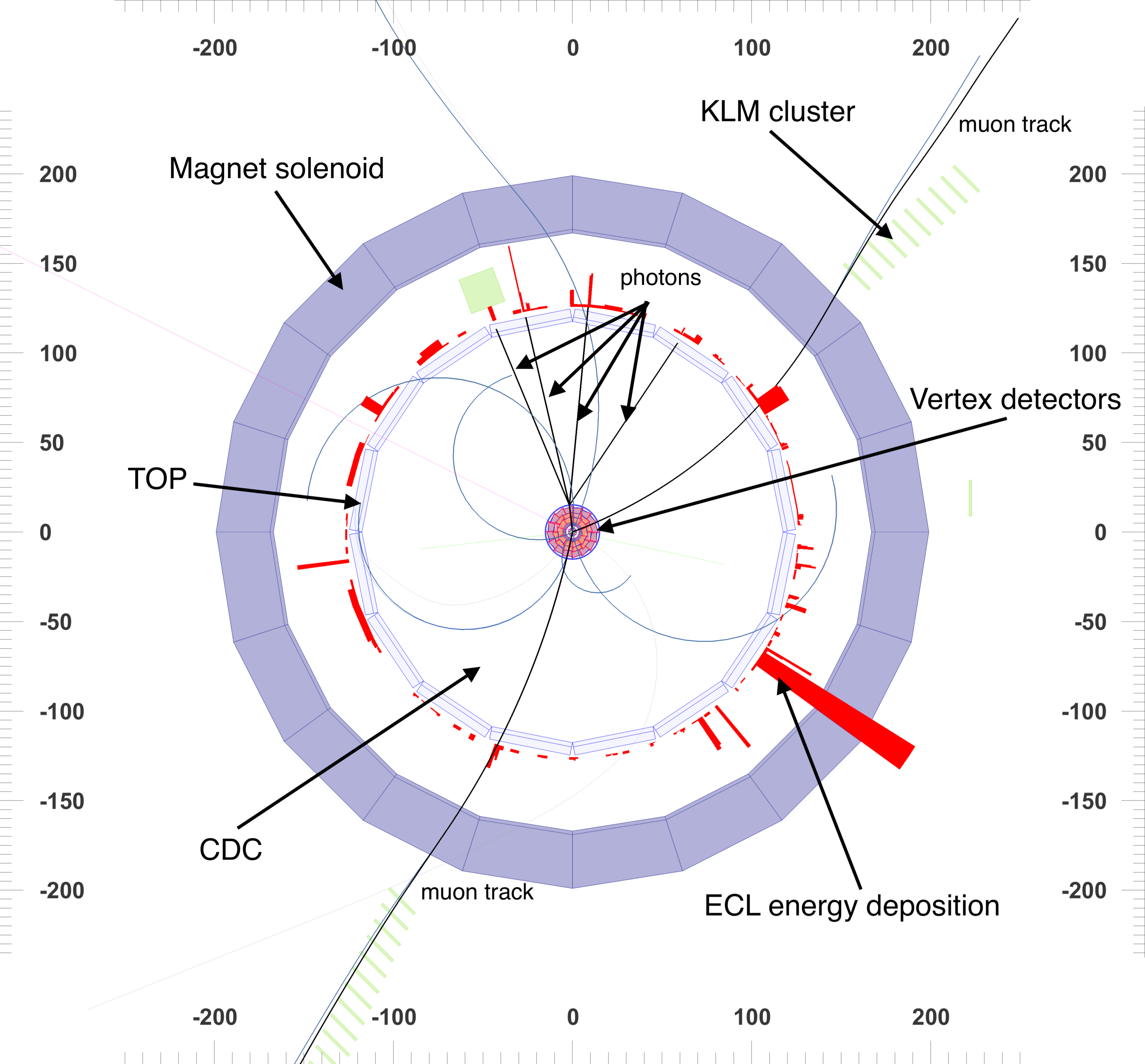}	
		\caption{ Event display projected onto the $x-y$ plane depicting the process of $ \bar{e}^{+}e^{-} \to \Upsilon(4{\rm{S}})\to \bar{B}^0 B^0$, with $\bar{B}^0 \to D\omega\pi\pi$ and $B^0 \to K_{\rm{S}} J/\Psi $, where $J/\Psi \to  \mu^+\mu^-$ and $ K^0_{\rm{S}} \to \pi^0 \pi^0$ with $\pi^0\to\gamma\gamma$. All particles of the signal $B^0$ decay are indicated by black lines. The rest of the event correspond to the decay of the $ \bar{B}^0$.  All distances are measured in centimetres. For a close-up of the vertex detector see Fig.~\ref{fig:zoom_in}.}
	\label{fig:zoom_out}
\end{figure}

\begin{figure}
		\centering
		\includegraphics[width=\linewidth]{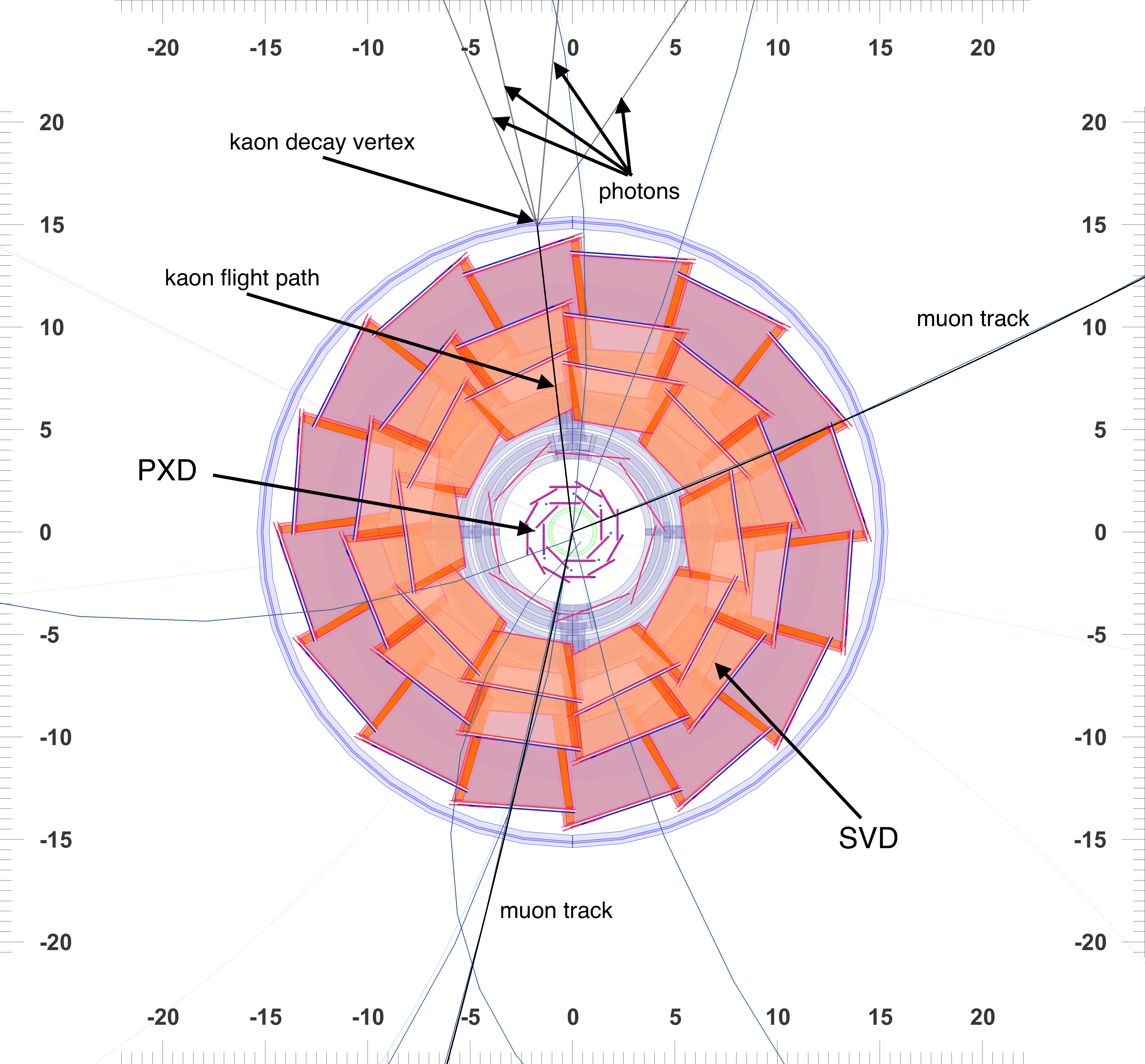}
	\caption{Close-up of Fig.~\ref{fig:zoom_out}. In this event, the$K^0_{\rm{S}}$ travels until the edge of the vertex detectors before decaying.}
		\label{fig:zoom_in}
\end{figure}

Fig.~\ref{fig:zoom_out} and Fig.~\ref{fig:zoom_in} show an event display depicting simulated particles traversing the Belle II detector. In the decay $\Upsilon(4{\rm{S}})\to \bar{B}^0 B^0$, one meson decays as $\bar{B}^0 \to D \omega\pi\pi$, and the other as $B^0 \to K_{\rm{S}} J/\Psi $, with $J/\Psi \to  \mu^+\mu^-$ and $K^0_{\rm{S}} \to \pi^0 \pi^0$ with $\pi^0\to\gamma\gamma$. Fig.~\ref{fig:zoom_out}, depicts the full detector geometry and Fig.~\ref{fig:zoom_in} shows a close-up of the inner vertex detectors. In this example we show that the decay vertex of the $K_{\rm{S}}^0 $ can be very displaced.


\externaldocument{constraints}

\section{Extended Kalman Filter}
Vertex fitting is a least squares minimization problem. The best least squares estimator (LSE) is the solution to this problem. The computational challenge in finding a LSE lies in matrix inversions, which naively scale as $\mathcal{O}(n^3)$, where $n$ is the dimension of the matrix. In a naive approach, this is equal to the number of parameters extracted in the fit. 
A Kalman Filter is an iterative approach to find the LSE by defining a series of constraints (knowledge of parameters from measurements and symmetries) on a hypothesis (a set of particle parameters in this case). The hypothesis has different states during the stages of the filtering process.  The filtering process is an iterative algorithm applying the constraints sequentially and updating the state with respect to the constraints. The sequence is repeated until convergence is reached or divergence is observed. In the case of convergence, the last state describes the best hypothesis for the parameters that can be found.
We use an Extended Kalman Filter in the gain matrix formulation \cite{kalman, Fruhwirth:178627, Hulsbergen:2005pu} in our algorithm. 
The state vector $\bm{x}$ holds the particle parameters to be optimized by the fit. The parameters depend on the type of particle. The most general parametrization takes the form
  \begin{equation}\label{eq:statevector}
  \bm{x} = \{ x_1, y_1, z_1, \theta_1,  p_{x1}, p_{y1}, p_{z1}, E_1, ..., x_n, y_n, z_n, \theta_n,  p_{xn}, p_{yn}, p_{zn}, E_n \},
  \end{equation}
  where vertex coordinates of the $i$-th particle are denoted as $\{ x_i, y_i, z_i \}$, its decay length is denoted as $\theta_i $ and the four momentum is $\{ p_{x,i}, p_{y,i}, p_{z,i}, E_i \}$, and $n$ is the number of particles in the fitted topology. This implies that $n \approx 8\times \text{number of particles}$.
  The problem is split into $k$ constraints given by measurements and other knowledge of parameters (see Sec.~\ref{sec:constraint} for the definitions of the measurement $\bm{m}$ and hypothesis $\bm{h}$ for each of the different constraints). The minimized $\chi^{2}$ can be expressed as a weighted sum of $k$ sets of equations, which are the constraints. For a given iteration, $\alpha$, of the Kalman Filter, one can write
  \begin{equation}\label{chi2kalman}
  \begin{aligned}
  \chi^{2}_\alpha =& (\bm{m}_1 - \bm{h}_1(\bm{x}^\alpha_0))^{T} \bm{V}_1^{-1} (\bm{m}_1 - \bm{h}_1(\bm{x}^\alpha_0)) +\ ...\ \\ +&  (\bm{m}_k - \bm{h}_k(\bm{x}^\alpha_{k-1}))^{T} \bm{V}_k^{-1} (\bm{m}_k - \bm{h}_k(\bm{x}^\alpha_{k-1})),
  \end{aligned}
  \end{equation}
  using the measurement covariance matrix $\bm{V}_k^{-1}$ transported to the $\bm{m} - \bm{h} $ system as a weight, such that the hypothesis of each constraint $k$ depends on the outcome of the previous constraint.\\
  Minimizing Eq.~\ref{chi2kalman} yields a rule to find a new state $\bm{x}_{k}^\alpha $ for the current iteration and constraint, that is 
 \begin{equation}
  \bm{x}_k^\alpha= \bm{x}_{k-1}^\alpha - \bm{K}_k^\alpha \bm{r}_k^\alpha ,
  \end{equation}
  where $\bm{K}$ is a gain matrix, which will be defined below. However, we first define the residual $r$ as the distance between measurement and hypothesis 
  \begin{equation}\label{eq:res}
  \bm{r}_k^{\alpha} = \bm{m}_{k}  -  \bm{h}_k( \bm{x}^{\alpha}) \ .
  \end{equation}
 The hypothesis $\bm{h}_k( \bm{x}^{\alpha})$  can be linearised around a reference state of the previous iteration $\bm{x}^{\alpha-1} $ using the Jacobian 
 \begin{equation}\label{eq:jacobian}
 \bm{H}^{\alpha-1}_{k} = \frac{  \partial \bm{h}_k(\bm{x}^{\alpha-1})  }{   \partial \bm{x}^{\alpha-1}    } ,
 \end{equation}
 such that
  \begin{equation}\label{eq:linearised}
  \bm{h}_{k}( \bm{x}^{\alpha}) = \bm{h}_k( \bm{x}^{\alpha-1})  -  \bm{H}_{k}^{\alpha-1} \cdot ( \bm{x}^{\alpha} - \bm{x}^{\alpha-1} ).
  \end{equation}
  Eq. \ref{eq:res} thus becomes
  \begin{equation}
  \bm{r}_{k}^{\alpha} =  \bm{m}_{k}  -  \bm{h}_k( \bm{x}^{\alpha-1})  +  \bm{H}_{k}^{\alpha-1} \cdot ( \bm{x}^{\alpha} - \bm{x}^{\alpha-1} ).
  \end{equation}
The gain matrix is calculated for every constraint in every iteration and is defined as
\begin{equation}
\bm{K}_{k} =  \bm{C}_{k-1} \bm{H}_{k}^T \bm{R}_k^{-1}.
\end{equation}
The only matrix to be inverted is of the dimension of the constraint and not the dimension of the statespace.
The current state's covariance matrix is obtained via propagation of uncertainties
\begin{equation}\label{eq:covupdate2}
\begin{aligned}
\bm{C}_k =&\ (1-{\bm{K}}_{k}{\bm{H}}_k) {\bm{C}}_{k-1} (1-\bm{K}_k\bm{H}_k)^{T}\\
=&\ \bm{C}_{k-1} - \bm{C}_{k-1} \bm{H}_{k}  \bm{R}^{-1}_k  \bm{H}_{k} \  \bm{C}_{k-1}^T \\
=&\ \bm{C}_{k-1} - \bm{C}_{k-1} \bm{H}_k^T  \bm{K}_{k}^T \ .
\end{aligned}
\end{equation}
The covariance matrix of the residual system can be found to be
\begin{equation}
\bm{R}_{k} = \bm{V}_{k} + \bm{H}_{k} \bm{C}_{k-1}\bm{H}_{k}^T.
\end{equation}
For constraints where no measurement is available, for example the mass constraint, we set $\bm{V}_{k}=0$.\\
Linearising the hypothesis around a reference state, see Eq.~\ref{eq:linearised}, makes the fits more stable than a normal Kalman Filter and reduces the overall number of signal candidates for which the fit diverges.
 In this formulation of the LSE problem, the update of the covariance matrix, Eq.~\ref{eq:covupdate2}, is the computational bottleneck, as it is a dense $m \times m$ matrix with the dimension of the state vector $m$ and must be calculated and filled $k$ times for each iteration.

\section{Parametrising and constraining the decay chain}\label{sec:constraint}

To parametrise the decay chain, we use a set of parameters that describe the properties of the particles. We perform a number of reductions on these properties to reduce the dimensionality of the problem.
For final state particles, we only save the momenta, as they do not have decay vertices. For their production vertices we use the decay vertices of their respective composite particles, later referred to as ``mothers". The energy of each each final state particle is calculated using the momenta and it's nominal mass hypotheses, taking the mass values provided by the particle data group (PDG) \cite{PDG}.
 Intermediate particles are classified in two categories: particles that decay dominantly via the strong force and via the weak force. Strongly decaying particles with a lifetime of less than $10^{-14}\e{s}$, which corresponds to boosted flight lengths of the order of less than $1\e{\mu m}$, are treated as if they instantly decay (the detector has a vertex position resolution in the x-y plane for charged particles of $20-30\e{\mu m}$). The hypothesised quantities are their energy and momenta, while the production and decay vertices coincide and are taken from their mother particle's decay vertex. For weakly decaying particles, we additionally measure a decay vertex and a flight length, defined as the distance between the production and decay vertices in three dimensions.
\subsection{Parametrising the constraints}
Constraints are defined by Eq.~\ref{eq:linearised}. The resulting Jacobians take the form of $m \times n$ matrices where $m$ is the dimension of the state vector and $n$ is the dimension of the respective constraint. Thus, only few of its elements are non-zero. For example, for a three dimensional point constraint $k$, the hypothesis of particle number $j$ with $\bm{h}_j = \{x_j,y_j,z_j\}$ and $\bm{x}$ as in Eq.~\ref{eq:statevector}, only the $j$-th diagonal block is non-zero
\begin{equation}
\frac{\partial \bm{h}}{\partial \bm{x} } = \bm{H}=
\left(
\begin{array}{ccccc}
&           \\
\mbox{\Large 0}& & \Large{\bm{\mathbbm{1}}}_3 & &\mbox{\Large 0} \\
&  &  &  &   \\
\end{array}
\right)~.
\end{equation}
The blocks filled with zero correspond to the parameters of particle $\bm{x}_i \neq \bm{x}_j$.
We will omit the columns filled with zeros throughout this section, for brevity.\\
In the following we list the definitions of constraints that have been implemented in the Belle II software, based on the specific geometry of Belle II.
\subsubsection{Reconstructed track}
\begin{figure}[h]
	\centering
	\includegraphics[width=\textwidth]{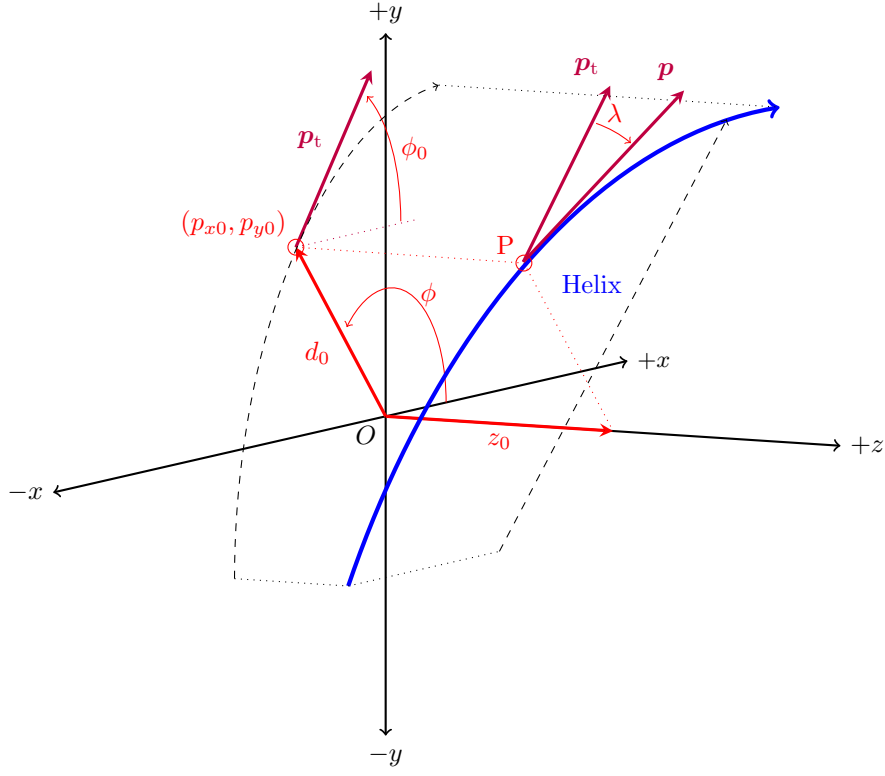}
	\caption{The perigee parametrisation of the track helix, depiction adapted from \cite{noel}. A description of the parameters can be found in Tab.~\ref{tab:helix}.}
	\label{fig:helix}
\end{figure}

A track can be parametrised with a five parameter helix. In Belle II it was chosen to use a perigee-parametrised helix, such that the helix is defined at the perigee, the point of closest approach of the helix to the origin of the coordinate system. The corresponding transformations to transport a helix to that point are discussed in Ref.~\cite{Karimaki}. A description of the parameters can be found in Table~\ref{tab:helix}, and a depiction of the helix is in Fig.~\ref{fig:helix}. We parametrise tracks such that we can express the model's dependence on Cartesian parameters as
\begin{equation}
\bm{h}_{\rm{track}}(\bm{x}) =
\begin{pmatrix}
d_{0 \rm{}}\\
\phi_{0 \rm{}} \\
\omega_{\rm{}} \\
z_{0 \rm{}} \\
\tan  \lambda_{\rm{}} \\
\end{pmatrix}
=
\begin{pmatrix}
A(1+U)^{-1}         \\
{\rm{atan2}}(p_{y}, p_{x}) - {\rm{atan2}}( \omega \cdot \Delta_{\parallel}, 1+\omega \cdot \Delta_{\perp} )  \\
a \cdot q / p_{t}\\
z + l \cdot \tan \lambda \\
p_z / p_{t} \\
\end{pmatrix}.
\end{equation}
Where atan2 refers to the phi domain corrected inverse tangent function.
We use the same parametrisation for the hypothesis and the measurement. We label the measurement quantities with the index $m$. The residuals of iteration $\alpha$ then become
\begin{equation}
\bm{r}_{{track}}^\alpha(\bm{x}) =
\begin{pmatrix}
d_{0, {m}} - d_{ 0}\\
\phi_{0, {m}} - \phi_{0} \\
\omega_{{m}} - \omega\\
z_{0, {m}} - z_{ 0}\\
\tan  \lambda_{{m}} - \tan \lambda\\
\end{pmatrix} + \bm{H}_{}^{\alpha-1} \cdot ( \bm{x}^{\alpha} - \bm{x}^{\alpha-1} ) ~.
\end{equation}
\begin{table}\centering
	\caption{Definitions of the Belle II helix parametrisation and their dependencies.}
	\label{tab:helix}
\begin{tabularx}{\textwidth}{c p{.4\textwidth} c  p{.6\textwidth} } 

	\hline

$d_{0}$ & The distance of closest approach to the $z$ axis (POCA)
signed with the $z$ component of the angular momentum w.r.t. to the origin.\\
$\phi_{0}$ &  Azimuthal angle of the momentum
at the POCA.\\
$\omega$ & Scaled inverse of the track momentum.  \\
$z_0$ &  The pivotal point is the perigee that is the POCA.\\
$\tan \lambda$ &  The angle of the momentum at the POCA with respect to the $x-y$ plane.\\
    \hline{}
    $\phi= {\rm{atan2}}(p_{y}, p_{x})$ & Angle of the $x-y$ plane to the helix.\\
    $\Delta_{\parallel} = - x\cdot \cos \phi - y\cdot \sin \phi$   & \multirow{4}{=}{Quantities used to transport the coordinate system, such that $\{x,y,z \}$ points to the perigee. }\\
    $\Delta_{\perp} = - y\cdot \cos \phi + x\cdot \sin \phi$   & &\\ 
    $A = 2 \cdot \Delta_{\perp} + \omega \cdot ( \Delta_{\perp}^2 + \Delta_{\parallel}^2 )$ & & \\
    $U = \sqrt{1+ \omega \cdot A}$& &\\
    $\lambda$ & Angle between the $x-y$ plane and the helix.\\
    $l = {\rm{atan2}}( \omega \cdot \Delta_{\parallel}, 1+\omega \cdot \Delta_{\perp} )$ & Arc length from the perigee to a point. \\
    $p_{x},~p_{y},~p_z$ & Momenta along the $x,~y,~z$ axes.\\
    $q$ & Charge of the particle.\\
    $B_z$ & Magnetic field strength in $z$-direction.\\
    $a = B_z / c$ & Magnetic field strength in the $z$ direction divided by the speed of light. \\ 

     $p_t = \sqrt{p_x^2 + p_y^2}$ & Transverse momentum.\\
      $p_{t0} = \sqrt{p_{x0}^2 + p_{y0}^2}$ & Pseudo transverse momentum.\\
       $p_{x0} = p_x - a\cdot q\cdot y $ & Pseudo momentum along the $x$ direction.\\
       $p_{y0} = p_y + a\cdot q\cdot x $  &Pseudo momentum along the $y$ direction. \\
       $r^2 = x^2 +y^2 $ & Radius squared.\\ 
       $\beta = 1 + \frac{p_{t0}}{p_t}$ & Quantity used for reading convenience. \\ 
        \hline
\end{tabularx}
\end{table}
We define the Jacobian block $\bm{A} := \partial \bm{h}/\partial \bm{x}$ as the derivatives with respect to the vertex position, and $\bm{B} := \partial \bm{h}/\partial \bm{p}$ as the derivatives with respect to momentum. The positions of these blocks in the Jacobian depend on the topology fitted and the particle represented by the track.
We choose to order the state vector hierarchically. This means that the decay vertex parameters come before its momentum, followed by the daughter particle's parameters. The full Jacobian $\bm{H}$ then takes the following form
\begin{equation}
\bm{H}=
\left(
\begin{array}{ccccccccc}
...& & &  & ... & & &  & ...   \\
...& & &  & ... & & &  & ...   \\
...&  & \bm{A} &  & ...&  & \bm{B}&  & ...\\
...& & &  &...& & &  &...\\
...& & &  & ... & & &  & ...   \\
\end{array}
\right)~.
\end{equation}
For the non-zero elements of the Jacobian blocks, denoted by $\partial d_0/\partial x=\bm{A}_{d_0, x}  $, we derive the spatial components as
\begin{equation}\label{eq:Jacobian_chaos1}
\begin{aligned}
 \bm{A}_{d_0, x} &= \frac{p_{y0}}{p_{t0}} ,& \bm{A}_{d_0, y} &= -\frac{p_{x0} }{p_{t0}},&  &\\
 \bm{A}_{\phi_0, x} &=  \frac{ a\cdot q \cdot p_{x0}  }{  p_{t0}^2  } ,& \bm{A}_{\phi_0, y} &=  \frac{ a\cdot q \cdot p_{y0}  }{  p_{t0}^2   },&\\
 \bm{A}_{z_0, x} &= - \frac{ p_{x} \cdot p_{x0} }{  p_{t0}^2  } ,& \bm{A}_{z_0, y} &=     - \frac{ p_{x} \cdot p_{y0}}{  p_{t0}^2  },& \bm{A}_{z_0, z} &= 1 ,   \\
\end{aligned}
\end{equation}
and for the momenta
\begin{equation}\label{eq:Jacobian_chaos1b}
\begin{aligned}
	&\  \bm{B}_{d_0, p_x} = \begin{aligned}[t] &\  -\frac{  y ((aq)^2  r + 2 aq  p_y  x + 2  p_y^2  \beta)	  }{  p_{t0} p_{t}^2 \beta^2  }\\
		&\  -\frac{ p_x (2  p_y  x \beta + aq  (y^2 (-2+ \beta ) + x^2 \beta )} {  p_{t0} p_{t}^2 \beta^2  },   \end{aligned}\\
	&\  \bm{B}_{d_0, p_y} =  \begin{aligned}[t]& \frac{ 2  p_x^2  x \beta + 2  p_x  y  (p_y - aq  x + p_y p_{t0}/p_t )     }{  p_{t0} p_{t}^2 \beta^2  } \\
		&+ \frac{aq (aq  r x - p_y  (x^2  (-2 +\beta) + y^2  \beta))}{  p_{t0} p_{t}^2 \beta^2  }, \end{aligned}\\
		\end{aligned}
\end{equation}
and
\begin{equation}\label{eq:Jacobian_chaos1c}
\begin{aligned}
\bm{B}_{\omega, p_x}\ &= -  \frac{aq p_x}{ p_t^3},\ \bm{B}_{\omega, p_y} = -  \frac{aq p_y}{ p_t^3},\\
\bm{B}_{z_0, p_x}\ &= \frac{p_z  (p_x^2  x - p_y  (aq  r + p_y  x) + 2  p_x  p_y  y)}{ p_{t0}^2 p_{t}^2},\\
\bm{B}_{z_0, p_y}\ &=  \frac{p_z  (p_x  (aq  r + 2  p_y  x) - p_x^2  y + p_y^2  y)}{p_{t0}^2 p_{t}^2},\\
\bm{B}_{z_0, p_y}\ &= (aq)^{-1} {\rm{atan2}}\left(aq ( p_y  y - p_x  x),\ p_x^2 + p_y  p_{y0} - aq  p_x  y\right),  \\
\bm{B}_{\tan \lambda, p_x} &= -  \frac{p_z p_x}{ p_t^3},\ \bm{B}_{\tan \lambda, p_y} = -  \frac{p_z p_y}{ p_t^3},\ \bm{B}_{\tan \lambda, p_z} =  p_t^{-1} .\\
\end{aligned}
\end{equation}

\subsubsection{Reconstructed photon}
\begin{figure}\centering

	\begin{subfigure}{.45\textwidth}
				\centering
		\begin{tikzpicture}[scale = .97]
\draw[-latex,line width=.8pt] (0,0)--(5,0) node[right]{$x$};
\draw[-latex,line width=.8pt] (0,0)--(0,5) node[above]{$y$};
\draw[line width=.8pt,black,-latex](0,0)--(2.5,1);

\node (A) at (2,0.25)  {$\boldsymbol{u}_{\rm{mother}}$};
\draw[line width=.8pt,black,-latex](2.5,1)--(3,4); 
\node (B) at (3.5,2)  {$\bm{\delta} = \tau \cdot \bm{p}_\gamma$};
\draw (2.5,1) node[circle, draw] {};

\draw[line width=.8pt,black,-latex](0,0)--(3,4); 
\node (C) at (1.5,2.5)      {$\boldsymbol{m}$};
\draw (3,4) node[circle,draw] {};

\draw (4.35,4.25)  node[anchor=south east] {Calorimeter cluster};
		\draw (5.75,1) node[anchor=south east] {Production vertex};
\end{tikzpicture}
\end{subfigure}
\centering
	\begin{subfigure}{.45\textwidth}

				\begin{tikzpicture}[scale = .97]
		\draw[-latex,line width=.8pt] (0,0)--(5,0) node[right]{$x$};
		\draw[-latex,line width=.8pt] (0,0)--(0,5) node[above]{$y$};

		\draw[line width=.8pt,black,-latex](0,0)--(2.5,1);
		\node (A) at (2,0.25)  {$\boldsymbol{u}_{\rm{mother}}$};
		
		\draw[line width=.8pt,black,-latex](2.5,1)--(3,4); 
		\node (B) at (3.7,2)  {$\bm{\Delta} = \theta \cdot \frac{\bm{p}}{|\bm{p}|}$};
		\draw (2.5,1) node[circle, draw] {};
		
		\draw[line width=.8pt,black,-latex](0,0)--(3,4); 
		\node (C) at (1.5,2.5)      {$\boldsymbol{u}$};
		\draw (3,4) node[circle,draw] {};
		
		\draw (4.1,4.25) node[anchor=south east] {Decay vertex};
		\draw (5.75,1) node[anchor=south east] {Production vertex};
		\end{tikzpicture}
	\end{subfigure}
		\caption{
	(a) The photon constraint, Eq.~\ref{eq:photo}, reduced to two dimensions for simplicity. The vector $\bm{\delta}$ is defined as pointing from the photon's production vertex to the measured calorimeter cluster, indicated with the photons mother's coordinate vector $\bm{u}$ and measurement vector $\bm{m}$. (b) Geometric constraint, Eq.~\ref{eq:geo1}. The vector $\bm{\Delta}$ is defined pointing from the particles decay and production vertex, indicated with the particle's and its mother's coordinate vector $\bm{u}$.  }
		\label{fig:photogeoconstr}
\end{figure}
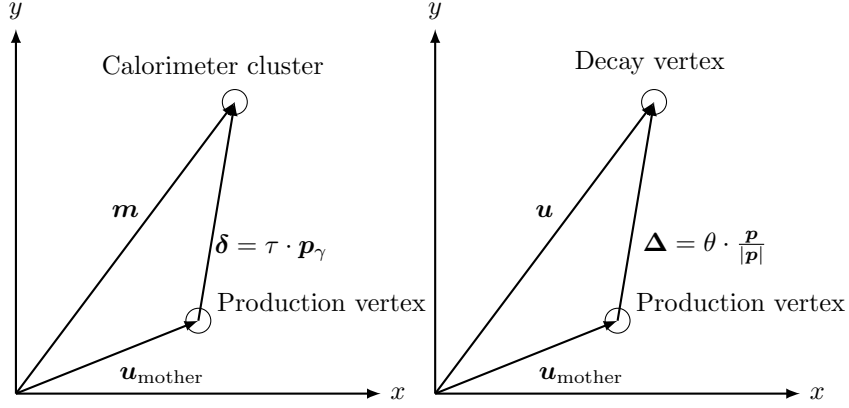
For photons we measure the position of the calorimeter cluster and its energy and can infer the vertex parameters. The geometry, depicted in Fig.~\ref{fig:photogeoconstr},  gives
\begin{equation} \label{eq:pthons}
  0 =  \bm{u}_{\rm{mother}} + \bm{\delta} -   \bm{m} ,
\end{equation}
substituting $\bm{\delta} = \tau \cdot \bm{p}$ and inserting the energy relation, we get
\begin{equation}
	\bm{h}_{{\rm{photon}}}(\bm{x}) =
	\begin{pmatrix}
	 u_{x} + \tau \cdot p_x \\
	 u_{y} + \tau \cdot p_y \\
	 u_{z} + \tau \cdot p_z \\
	 \sqrt{p_x^2+p_y^2+p_z^2}
	\end{pmatrix}\ \  {\rm{and}} \ \ \bm{m}_{{\rm{photon}}}(\bm{x}) =\
	\begin{pmatrix}
	m_{x} \\
	m_{y}  \\
	m_{z}  \\
	E_{m}
	\end{pmatrix},
\end{equation}
where $\{u_x,u_y,u_z\}$ are the production vertex coordinates, $\{p_x,p_y,p_z\} $ are the parameters of the momentum vector pointing from the production vertex to the calorimeter cluster, $\{ 	m_{{x}}, m_{{y}}, m_{{z}}, E_{{m}}  \}$ are the position and measured energy of the corresponding ECL cluster. The parameter $\tau$ is the decay time, which can be eliminated when writing down the residual to reduce the dimensionality of the equation system and avoid a trivial local minimum of $\bm{r}_{\gamma}$ at $\tau=0$ when taking $\{u_x,u_y,u_z\}=0$ as the starting point of the first iteration.
Since the geometry of the detector is cylindrical, we can not simply eliminate any of the dimensions as this could introduce a pole in the residual equations. Thus, we sort the momenta and eliminate the dimension with the highest momentum. Such that we get a 3-dimensional equation system
\begin{equation}\label{eq:photo}
\bm{r}^{\prime \alpha}_{{\rm{photon}}}(\bm{x}) =
\begin{pmatrix}
(m_{i}- u_{i})  - (m_{k} - u_{k}) \frac{ p_{i}}{p_{k}} \\
(m_{j} - u_{j})  - (m_{k} - u_{k}) \frac{ p_{j}}{p_{k}} \\
E_{m} - \sqrt{p_{{i}}^2+p_{{j}}^2+p_{{k}}^2}
\end{pmatrix}  + \bm{H}_{}^{\alpha-1} \cdot ( \bm{x}^{\alpha} - \bm{x}^{\alpha-1} ) ~,
\end{equation}
where the indices $i,j,k$ indicate the dimensions by order of increasing momentum $p_k\ge p_i \ge p_j$. We define $\bm{A}_{i,u_k} := \partial h'_i/\partial u_k$ and $\bm{B}_{i,p_k} := \partial h'_i/\partial p_k $ with the hypothesis of the reduced system $r'$. Thus, the non-zero entries are
\begin{equation}\label{eq:Jacobian_chaos}
\begin{aligned}
\bm{A}_{0,u_k} &= \frac{p_i }{  p_k },& \bm{A}_{0,u_i} &= -1, & &\\
\bm{A}_{1,u_k} &= \frac{p_j }{  p_k },& \bm{A}_{1,u_j} &= -1, & &\\
\bm{B}_{0,p_k} &= p_k^{-2} ,& \bm{B}_{0,p_i} &= \frac{  u_{{k}} - m_{{k}} }{p_{{k}}} ,& &\\
\bm{B}_{1,p_k} &= p_k^{-2} ,& \bm{B}_{1,p_j} &= \frac{ u_{{k}} - m_{{k}}  }{p_{{k}}},& &\\
\bm{B}_{2,p_k} &= - \frac{p_k}{|\bm{p}|} ,& \bm{B}_{2,p_i} &= -\frac{p_i}{|\bm{p}|} ,& \bm{B}_{2,p_j} &= -\frac{p_j}{|\bm{p}|}. \\
\end{aligned}
\end{equation}
The full Jacobian then takes the form
\begin{equation}
\bm{H}=
\left(
\begin{array}{ccccccccc}
...& & &  & ... & & &  & ...   \\
...&  & \bm{A} &  & ...&  & \bm{B}&  & ...\\
...& & &  &...& & &  &...\\
\end{array}
\right)~.
\end{equation}
We must transform the covariance matrix of the measurement into the reduced system. For that, we use
\begin{equation}
\bm{V'}= \bm{F} \bm{V} \bm{F}^T,
\end{equation}
with the transport matrix $\bm{F} = \partial r' /\partial m $, which depends on the sorting of the momenta such that the non-zero entries are
\begin{equation}
\begin{aligned}
\bm{F}_{0,m_k} &= -\frac{p_i}{p_k},& \bm{F}_{0,m_i} &= 1,~ \\
\bm{F}_{1,m_k} &= -\frac{p_j}{p_k},& \bm{F}_{1,m_j} &= 1, \\
\bm{F}_{2,E_k} &= 1 .& & \\
\end{aligned}
\end{equation}
We do not parametrise this constraint in $p_t$ in order to keep the derivatives in Eq.~\ref{eq:Jacobian_chaos} as computationally simple as possible.
\subsubsection{Reconstructed  $K^0_L$ }
We treat $K^0_L$ in the same way as photons, except that we use the nominal mass provided by the PDG in the energy calculation. The KLM detector is used for the cluster position measurement instead of the calorimeter. It cannot provide a precise energy measurement. Instead, it extrapolates the energy deposited by a particle as $E = c \cdot n$, where $n$ is the number of hit cells in the cluster and $c$ is a constant with the units $\ee{GeV}$. This approach makes the energy measurement for $K^0_L$ much less resolved than for photons.
\subsubsection{Kinematic constraint}
The kinematic constraint enforces four-momentum conservation, meaning it fits the four-momentum of the mother as the sum of the daughter momenta
\begin{equation} \label{eq:kine}
\bm{r}^\alpha(\bm{x}) = \bm{p}_{\rm{particle}} - \sum_i \bm{p}_{i,\rm{daughter}} + \bm{H}^{\alpha-1} \cdot ( \bm{x}^{\alpha} - \bm{x}^{\alpha-1} )~.
\end{equation}
The Jacobian for a particle with $two$ daughters, which in this example are taken to be stable particles, can be defined 
\begin{equation}
\bm{H}=
\left(
\begin{array}{ccccccccccccc}
...& & &  & ... & & &  & ... & & &  & ...  \\
...& & &  & ...& & &  & ...& & &  & ... \\
...&  & \bm{A} &  & ...&  & \bm{B}_1 &  & ...&  & \bm{B}_2 &  & ...\\
...& & &  &...& & &  &...& & &  &...\\
\end{array}
\right),
\end{equation}
with the blocks $\bm{A}, \bm{B} $ as
\begin{equation}
\bm{A} = \frac{\partial \bm{h}}{\partial \bm{p}_{{\rm{particle}} } } =\bm{\mathbbm{1}}_4,~
\end{equation}
and
\begin{equation}\label{eq:stable_particle_B}
\bm{B}_i = \frac{\partial \bm{h} }{\partial \bm{p}_{{\rm{daughter}},i  } } = -1\left(
\begin{array}{cccc}
1     & &&  \\
& 1 &  &\\
&  & 1 &   \\
p_{x,i}/E_i & p_{y,i}/E_i & p_{z,i}/E_i & 0   \\
\end{array}
\right),
\end{equation}
Note that the energy row of $\bm{B}_i$ depends on how the particle is parametrised, composite particles, for example, are parametrised with an energy variable in the state vector, resulting in $\bm{B} = - \bm{\mathbbm{1}}_4$, while for stable particles Eq.~\ref{eq:stable_particle_B} is used.

\subsubsection{Geometric constraint}
The geometric constraint fits the decay length parameter $\theta$ for composite particles, see Fig.~\ref{fig:photogeoconstr}. Accounting for the geometry we have
\begin{equation} \label{eq:geo}
0 = \bm{u}_{\rm{mother}}  +\bm{\Delta}  - \bm{u} ~.
\end{equation}
Instead of directly extracting a flight vector $\bm \Delta$, we use the unit vector of the momentum as it is well constrained by the previously filtered kinematic constraints, substituting  $\bm{\Delta}= \theta \cdot \bm{p} / |\bm{p}|$,  allows for a more accurate estimation of $\theta$. Thus we define the residual as
\begin{equation} \label{eq:geo1}
\bm{r}^\alpha(\bm{x}) = \bm{u}_{\rm{mother}}  + \theta \cdot \frac{\bm{p}}{|\bm{p}|}  - \bm{u} + \bm{H}^{\alpha-1} \cdot ( \bm{x}^{\alpha} - \bm{x}^{\alpha-1} )~.
\end{equation}
using 
\begin{equation}
\bm{A} = \frac{\partial \bm{h}    }{\partial \bm{u}_{{\rm{mother}}  } } =  {\bm{\mathbbm{1}}}_3,~ 
\bm{B} = \frac{\partial \bm{h}    }{\partial \bm{u} }= -{\bm{\mathbbm{1}}}_3  ,~
\bm{C}= \frac{\partial \bm{h}   }{\partial \theta } = \frac{1}{|\bm{p}|}  \left(
\begin{array}{c}
p_x \\
p_y  \\
p_z \\
\end{array}
\right),~
\end{equation}
and
\begin{equation}
\bm{D} = \frac{\partial \bm{h}    }{\partial \bm{p} }= \frac{\theta}{|\bm{p}|^{3}} \left(
\begin{array}{ccc}
( p_y^2 + p_z^2  )    & -p_x p_y &  -p_x p_z\\
-p_y p_x& ( p_x^2 + p_z^2  )  & -p_y p_z \\
-p_z p_x&  -p_z p_y& ( p_x^2 + p_y^2  )    \\
\end{array}
\right),
\end{equation}
such that
\begin{equation}
\bm{H}=
\left(
\begin{array}{ccccccccccccccccc}
...&&   &   & ...& &    &     & ... &   &    &    & ...&   &    &    & ... \\
...& & \bm{A} &   & ...&   & \bm{B}&     & ... &   &\bm{C}&     & ...&      & \bm{D}   &    & ...\\
...&   &  & & ...&   &    & & ... &    &     & & ...&       &    & & ....\\
\end{array}
\right)\ .
\end{equation}

\subsubsection{Mass constraint}
The mass constraint requires a particle four-vector to be consistent with its nominal mass.
We treat the particle as a measurement with infinite precision and use the mass value provided by PDG such that
\begin{equation} \label{eq:massconstr}
r^\alpha(\bm{x})= m_{\rm{PDG}}^2 - E^2 + |\bm{p}|^2 - \bm{H}^{\alpha-1} \cdot ( \bm{x}^{\alpha} - \bm{x}^{\alpha-1} ),
\end{equation}
with
\begin{equation}
\bm{H}=
\left(
\begin{array}{cccccc}
...& 2p_x     & & &    & ...  \\
...& & 2p_y & &  & ... \\
...&   &  & 2p_z &  & ...  \\
...&  &  &   & -2E   & ...  \\
\end{array}
\right).
\end{equation}

\subsubsection{Beam spot constraint}

$B$-mesons have a short lifetime and decay very close the beam spot, therefore it is useful to constrain the reconstructed particles to a volume within that region.
The constraint is implemented by considering the initial $e^+e^-$ collision as an abstract mother particle with the parameters and uncertainties of the beam spot, thus constraining the $production$ vertex position of its daughters to an area within the beam spot's three dimensional uncertainty region using
\begin{equation} \label{eq:beamstr}
	\bm{r}^\alpha(\bm{x})= \bm{s} - \bm{h}_{\rm{production}}  + \bm{H}^{\alpha-1} \cdot ( \bm{x}^{\alpha}- \bm{x}^{\alpha-1} ),
\end{equation}
with the Jacobian
\begin{equation}
\bm{H}=
\left(
\begin{array}{ccccc}
...&   & &  & ...        \\
...& & \Large{\bm{\mathbbm{1}}}_3&   &...\\
...&  &  &  & ...  \\
\end{array}
\right).
\end{equation}
Here $\bm{s}$ denotes the beam spot position vector and  $\bm{h}_{\rm{production}}$ is the fitted production vertex of the particle. The beam spot is determined  as part of the detector calibration, by averaging the $x$, $y$ and $z$ positions of the $e^+e^-$ collisions over many interactions. The $decay$ vertex can  be constrained similarly.

\subsubsection{Custom Origin constraint}
Similar to the beam spot constraint, one can construct another geometric constraint by defining a custom vertex position and an associated uncertainty to be the origin of the decay chain. This can be very useful when the decay chain contains particles that can not be detected, for example, neutrinos or long lived dark matter particles. In these decay chains it is not useful to fit the full chain, because the missing particle's four momentum makes the kinematic constraint of the mother particle an incorrect assumption. However, knowing that the particle originates from a $B$-meson decay, one can define a geometric constraint corresponding to the volume where $B$-mesons decay on average.

A beam energy constraint is not needed, as one can mass constrain the $\Upsilon(4S)$ particle, since all four-vectors in $e^+ e^{-} \to \Upsilon(4\rm{S})$ are well known.

\section{Fitting complex decay topologies and extracting parameters from the fit}

In this section we present two use cases of the algorithm containing one and two $\pi^0$-mesons in a decay chain, respectively. There are numerous channels where the phase space is large enough to add $\pi^0$ to a vertex with two charged tracks, which makes this a very common structure in decay trees and therefore an interesting target to fit in a wide spectrum of analyses. We use Monte Carlo samples from the Belle II experiment and evaluate background rejection and signal resolution improvements, as well as the extraction of various other parameters. The selection criteria used for all analyses in this section are listed in Table~\ref{tab:precuts}. We use implicit charge conjugation, meaning that charged particles appearing in the decay chains imply including the opposite charges. We abbreviate decays $ x \to yz$ as $x(yz)$.

\subsection{Fitting decay chains containing a neutral particle}
\begin{figure}	\centering
	\begin{subfigure}{0.49\textwidth}
	\centering
	\includegraphics[width=\textwidth]{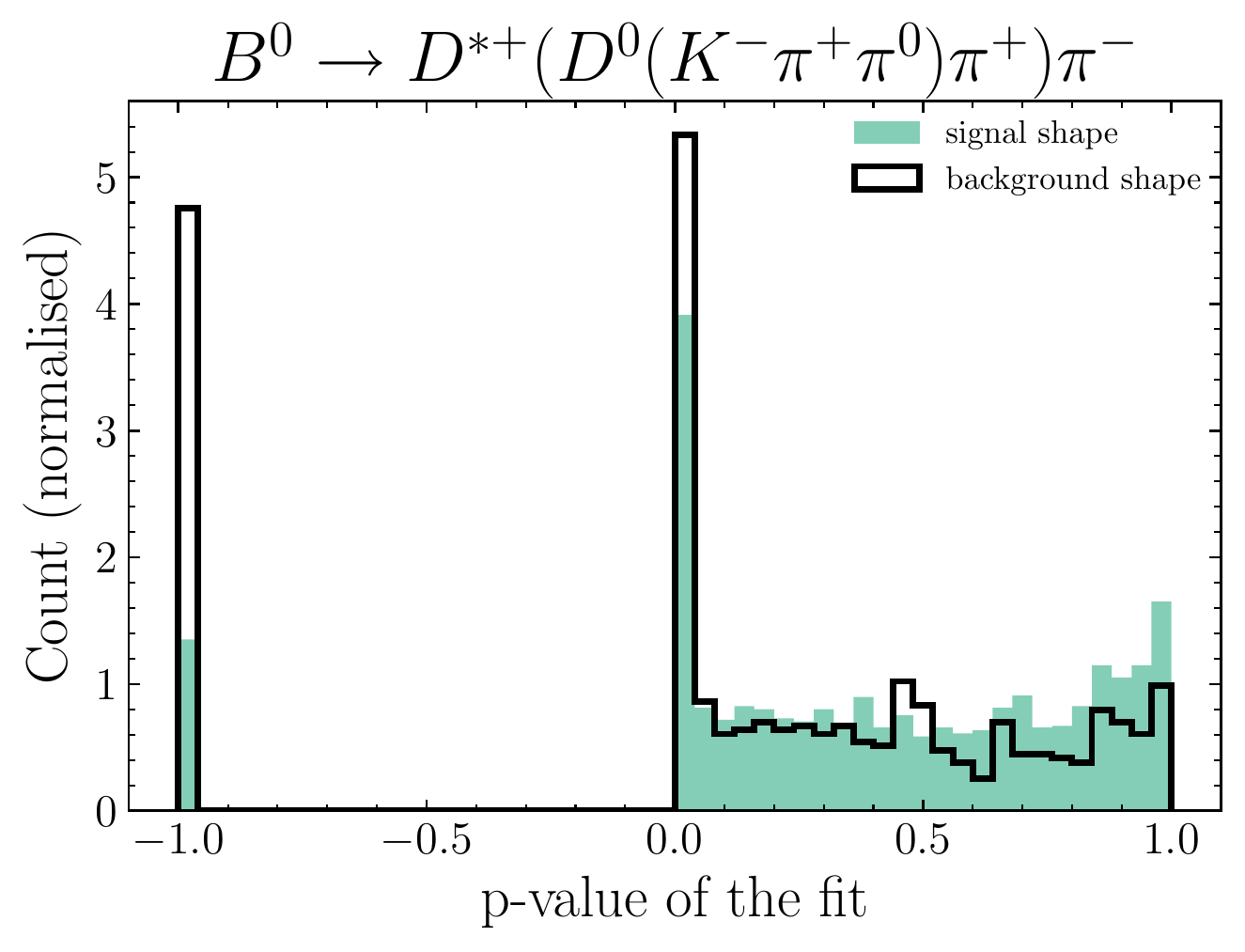}
\caption{   }
\label{fig:BdsDokPiPi0}
		\end{subfigure}
	\caption{P-value distributions of fits to $B^0\to D^{*+}(D^0(K^-\pi^+\pi^0) \pi^+) \pi^-$. The background and signal rejection properties of the p-value are insensitive to the number of DOF. The preselection criteria are listed in Table~\ref{tab:precuts}.  }
\end{figure}

The first mode, with a single $\pi^0$, is $B^0 \to D^{* +}({D}^0(K^- \pi^+\pi^0)\pi^+)\pi^-$. This channel is of interest as it is the most common $D^0$ decay modes.  We use the fitter in this example to suppress background. If we reject combinations for which the fit failed - i.e. those with a p-value $\le 0$ - we are able to reject about $20 \%$ of the background while rejecting only about $4\%$ signal, as displayed in Fig.~\ref{fig:BdsDokPiPi0}. Note that since some of the constraints are nonlinear, the $\chi^2$ which is used to calculate the p-values can be degenerate, where the effective number of degrees of freedom is smaller than the theoretical one used to calculate the p-values. As a result, the p-values can have non-flat shapes \cite{chi2}.

\subsection{Fitting ill-defined decay vertices}
\begin{figure}	\centering

		\begin{subfigure}{.49\textwidth}
	\centering
\includegraphics[width=\textwidth]{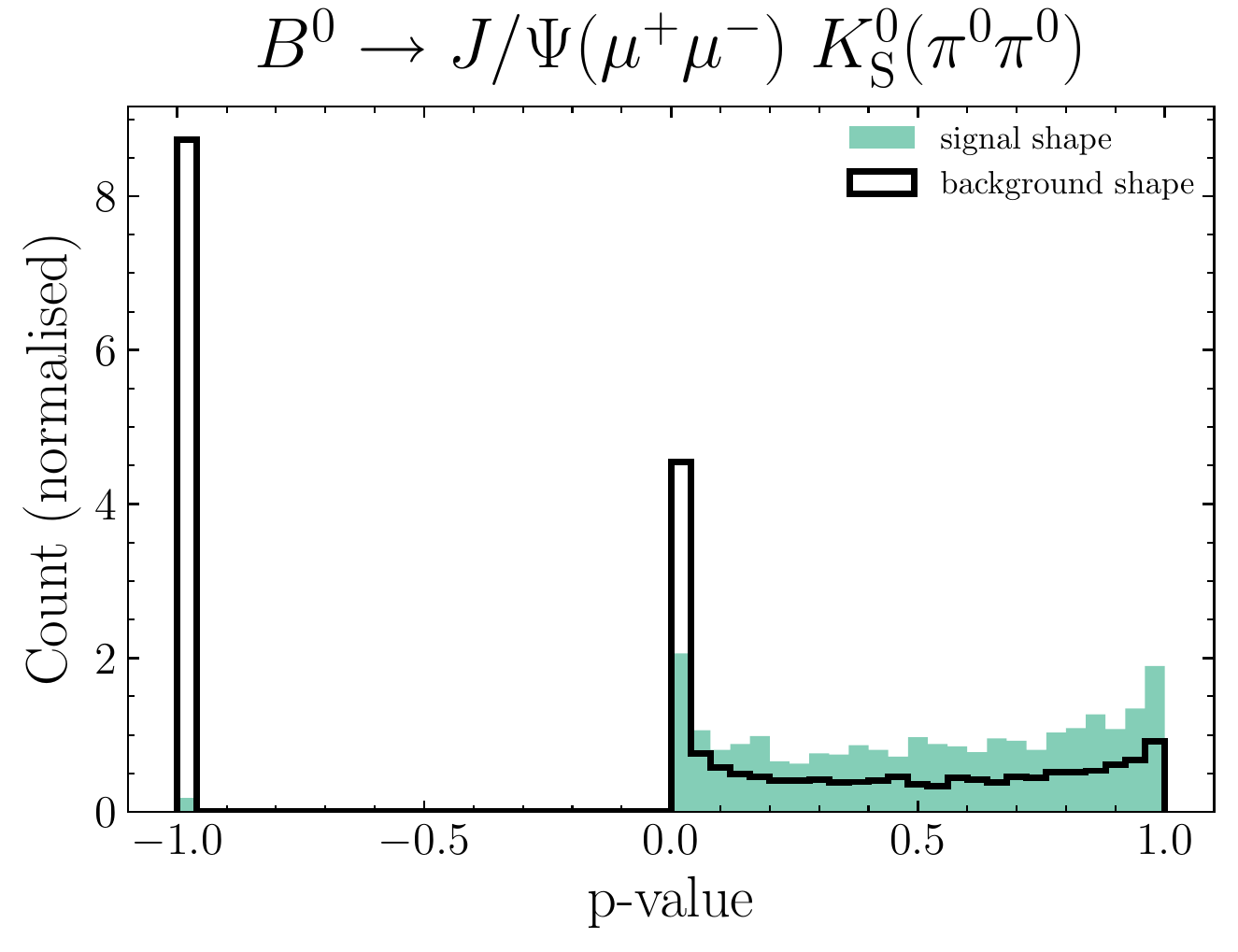}
\caption{   }
\label{fig:Ks_pval}
		\end{subfigure}
\begin{subfigure}{.49\textwidth}	\centering
		\centering
	\includegraphics[width=\textwidth]{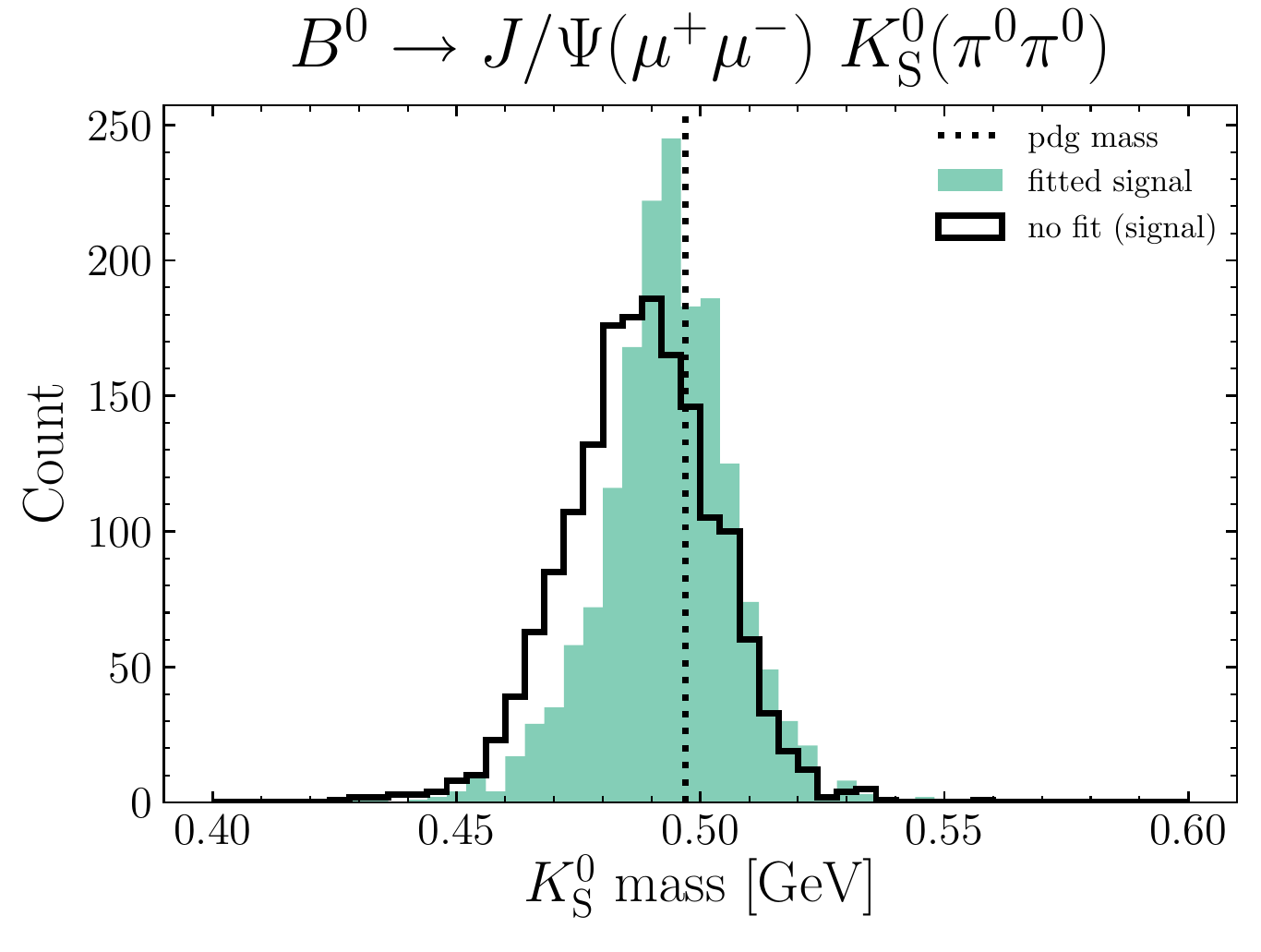}
\caption{   }
	\label{fig:Ks_mass}
	
		\end{subfigure}
		\caption{a) P-value of the fits to $B^0\to J/\psi K^0_{\rm{S}}$. b) Fitted mass of the $K^0_{\rm{S}}$ (green) and the mass before the fit (black). The mass distribution is centred around the true value after performing the fits with a $\pi^0$ mass constraint. The resolution has slightly improved as well.   }
\end{figure}

We study the decay chain $B^0 \to K^0_{\rm{S}} (\pi^0(\gamma\gamma)\pi^0( \gamma\gamma)) J/\Psi( \mu^+\mu^-)$. By performing the fit a large amount of the background can be removed by requiring that it passes the fit, see Fig.~\ref{fig:Ks_pval}. 
The only well defined vertex in this chain is given by the $J/\Psi \to \mu^+\mu^-$ decay. The other vertices are ill-defined, the best assumption that can be made for the production vertex position of the four photons is that they originate from the interaction point. This assumption biases the reconstructed mass of the $K^0_{\rm{S}}$.
Performing a fit with mass constrained $\pi^0$-mesons improves the extracted mass of the $K^0_{\rm{S}}$, so that after the fit, it is centred around the true value, as depicted in Fig.~\ref{fig:Ks_mass}. It is then possible to further reject background outside the nominal mass window and improve the signal purity when analysing this channel.

\subsection{Extracting the decay length of $D^0$-mesons from $D^{* +}$ decays using a beam-spot constraint}
\begin{figure}\centering
	\begin{subfigure}{.49\textwidth}
		\includegraphics[width=\linewidth]{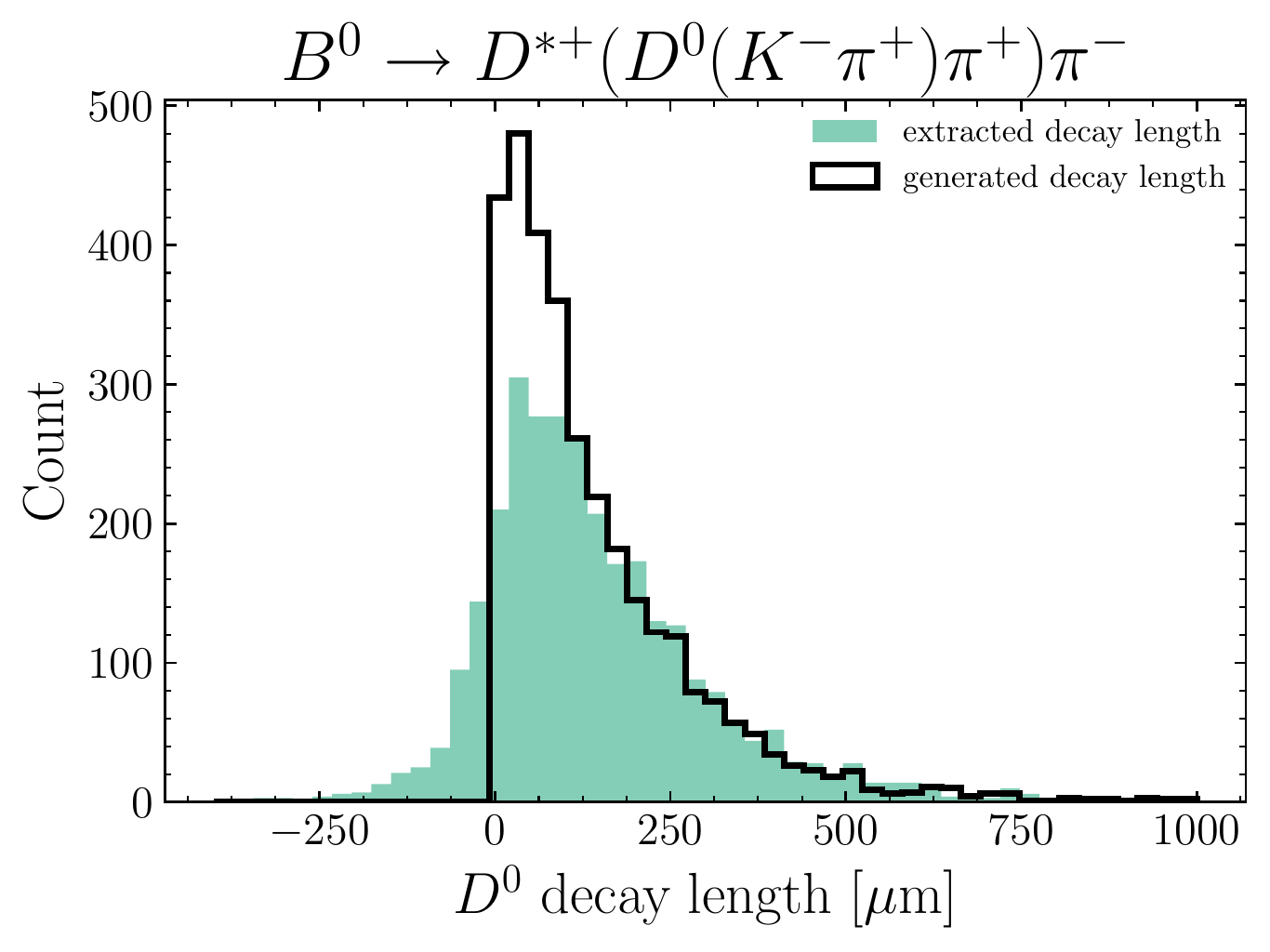}
		\caption{ }
		\label{fig:decaylength}
	\end{subfigure}
	\begin{subfigure}{.49\textwidth}
		\includegraphics[width=\linewidth]{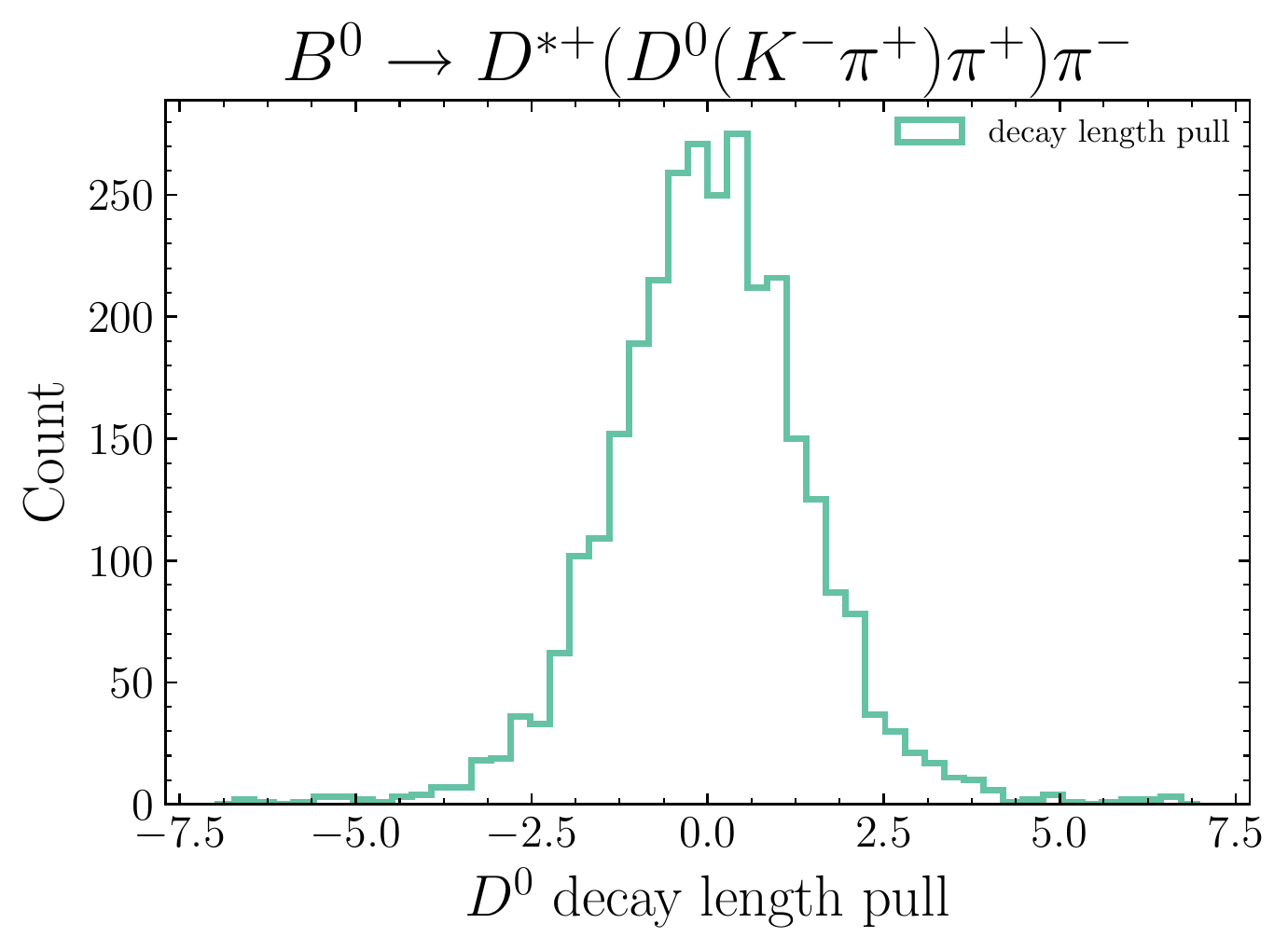}
		\caption{ }
		\label{fig:decaypull}
	\end{subfigure}
	\caption{ 
	a) The extracted decay length (green) and the generated decay length (black). The negative tail of the extracted decay length is due to the detectors resolution function.
	b) Pull ($ \rm{measured - generated / uncertainty } $) of the decay length distribution. 
					}
					\label{fig:decaylengtha}
\end{figure}

The geometric constraints allow for the extraction of all the production and decay vertices of all particles in the decay chain. This allows for the extraction of flight lengths and thus lifetimes of intermediate particles such as $D^0$-mesons. We perform this study on $B^0 \to D^{* +}( {D}^0( K^- \pi^+)\pi^+)\pi^-$ decays and extract the decay length of the $D^0$-meson. The results are depicted in Fig.~\ref{fig:decaylengtha}.
Since $D^{* +}$-mesons decay almost instantly, we will use the three dimensional distance between the $B^0$ and the $D^0$ decay vertices. To improve the resolution on the $B^0$ vertex we apply a beam-spot constraint. 

\subsection{Using a custom origin constraint to improve the $D^{* +}$ vertex resolution in $B^0 \to D^{* +}({D}^0( K^- \pi^+)\pi^+)l^-\bar{\nu}$ decays}
\begin{figure}
	\centering
	\begin{subfigure}{.49\textwidth}
		\includegraphics[width=\linewidth]{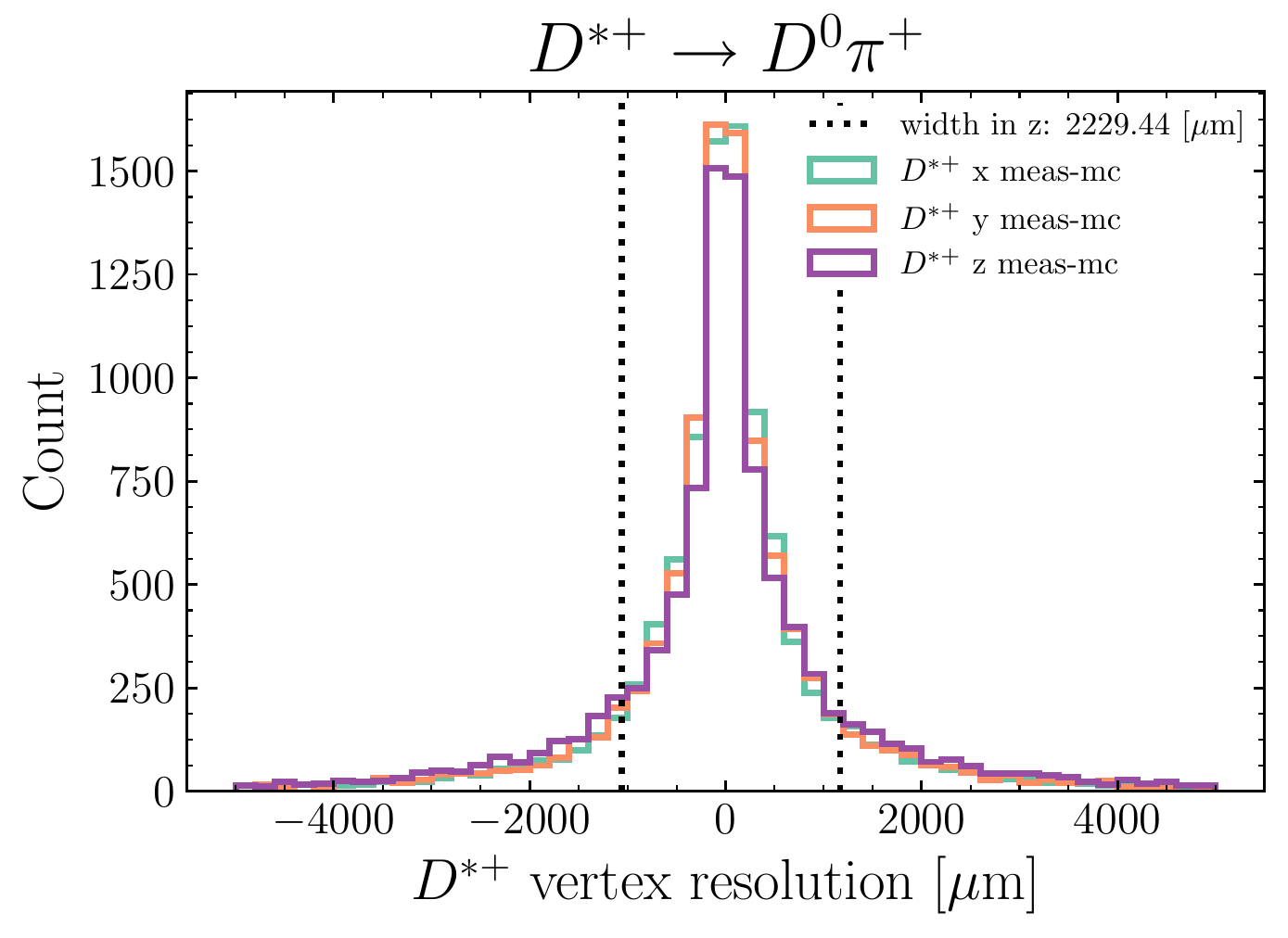}
		\caption{ }
		\label{fig:custom_origin1}
	\end{subfigure}
	\begin{subfigure}{.49\textwidth}
		\includegraphics[width=\linewidth]{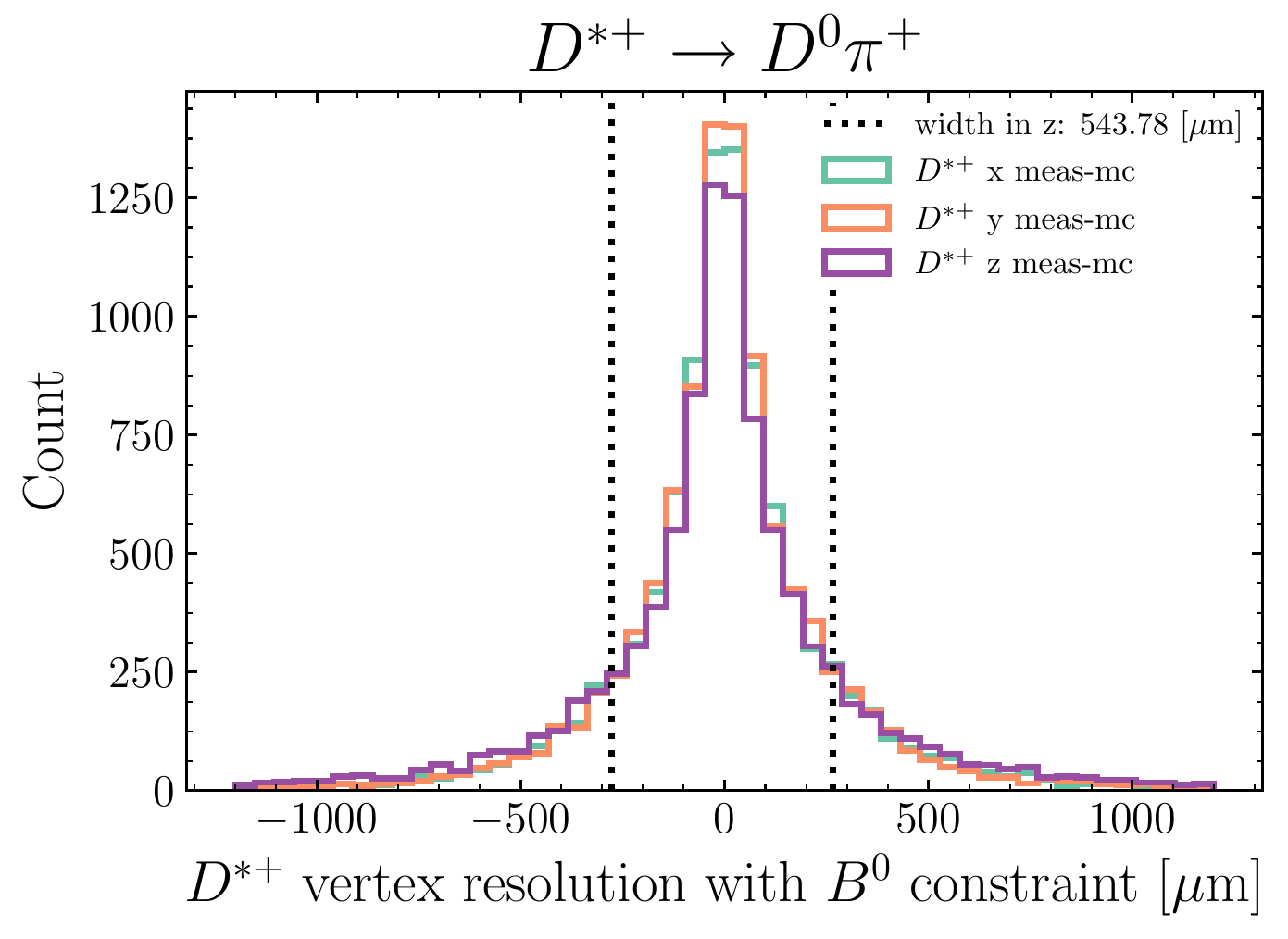}
		\caption{ }
		\label{fig:custom_origin2}
	\end{subfigure}
	\caption{Resolution (= measured - generated)  of the $D^{* +}$ decay vertex position extraction without (a) and with (b) a custom origin constraint. The vertex position and uncertainties for the origin used in this example are extracted from $B^0$ decays. }
		\label{fig:custom_origin}
\end{figure}

In some scenarios one can not fit the entire decay chain e.g. in semi-leptonic decays $B^0 \to D^{* } \mu \nu_\mu$, where the neutrino escapes detection. Since the neutrino's four vector is unknown it is not possible to use a kinematic constraint to fit the $B$-meson.
To circumvent this problem, one can omit fitting the $B^0$, and instead account for its presence with a geometric constraint. In $B^0 \to D^{* +}({D}^0( K^- \pi^+)\pi^+)l^-\bar{\nu}$ decays we use a custom origin constraint to constrain the $D^{*+}$ production vertices to a volume around the median $B$-meson decay vertex position obtained from Monte Carlo simulations. The results are depicted in Fig.~\ref{fig:custom_origin}. This technique can improve the resolution on the $D^{* +}$ vertex position. 

\begin{table}\centering
	\caption{Selection criteria used in the analyses in this section. All particles use the same criteria if appearing in the decay chains. The invariant mass is obtained by summing the particle's daughter four-momenta before performing the fit. The particle ID for $K^+/\pi^+$ is defined as the likelihood ratio $\mathcal{LR}(K^+(\pi^+)) = \mathcal{L}(K^+(\pi^+))/(\mathcal{L}(\pi^+) + \mathcal{L}(K^+))$ and the beam energy constrained mass $m_{{\rm{bc}}}$ \cite{b2book}.}
	\label{tab:precuts}
	\begin{tabular}{c l  c l} 
		particle & pre selection applied\\
		\hline{}
		$\gamma$ & $E_\gamma > 0.075\e{GeV}$ \\
		$\pi^0$ & $ 0.145\e{GeV} > m(\gamma\gamma) > 0.125\e{GeV} $ \\
		$\pi^+$ & $\mathcal{LR}(\pi^+)> 0.5$ \\
		$K^+$ & $\mathcal{LR}(K^+)> 0.5$ \\
		$D^0$ & $ 1.9\e{GeV} > m(K^-\pi^+\pi^0) > 1.7\e{GeV}$  \\
		$D^{*}$ & $ \Delta m(D^{*}-D^{0})  < 0.155\e{GeV} $  \\
		$B^0$ & $m_{{\rm{bc}}} = \sqrt{E^2_{{\rm{beam}}}/4 - p^2} > 5.27\e{GeV}$  \\
		\hline{}
	\end{tabular}
\end{table}

\section{Conclusion}
We presented a versatile fitting tool tailored for the environment of $B$~factories with a cylindrical detector geometry. It can be used for various purposes, such as the extraction of particle production and decay vertices, decay lengths, particle four-momenta and rejection of backgrounds, as well as the extraction of the respective uncertainties for all parameters involved in three dimensions. This global fitting technique is particularly powerful in fitting and reducing background in modes that contain neutral particles, and will be very important for the Belle~II physics analysis program. As a future extension of the fitter we are looking into the possibility to map the residuals found by the Kalman Filter to a scalar signal probability $f:\bm{r} \to p$. Such a probability may outperform methods using a p-value derived from the fit $\chi^2$ for the purpose of signal-background-separation.

\section{Acknowledgements}
This work was made possible by the Belle II collaboration and funding from ARC (Australia), ARRS (Slovenia), BMBF (Germany), EXC153 (Germany), HGF (Germany), INFN (Italy), MSMT (Czech Republic), NSERC (Canada) and U.S. DOE. We would like to thank Wouter Hulsbergen for his original paper, the discussion and sharing his code.

\nolinenumbers
\newpage
\bibliography{bibResources}

\end{document}